\documentclass[aps,prd,twocolumn,groupedaddress]{revtex4-1}

\usepackage{graphicx}
\begin{document}



\title{ARCHIMEDEAN-TYPE FORCE IN A COSMIC DARK FLUID: \\
II. QUALITATIVE AND NUMERICAL STUDY \\ OF A MULTISTAGE UNIVERSE EXPANSION}


\author{Alexander B. Balakin\footnote{email: Alexander.Balakin@ksu.ru}
and Vladimir V. Bochkarev\footnote{email: Vladimir. Bochkarev@ksu.ru}}
\affiliation{Kazan Federal University, Kazan, Russia}

\date{\today}

\begin{abstract}
In this (second) part of the work we present the results of  numerical and qualitative
analysis, based on a new model of the Archimedean-type interaction between dark matter and dark energy.
The Archimedean-type force is linear in the four-gradient of the dark energy pressure and plays a role of self-regulator of the energy
redistribution in a cosmic dark fluid. Because of the Archimedean-type interaction
the cosmological evolution is shown to have a multistage
character. Depending on the choice of the values of the model guiding parameters,
the Universe's expansion is shown to be perpetually accelerated, periodic or quasiperiodic with finite number of
deceleration/acceleration epochs. We distinguished the models, which can be definitely characterized by the inflation in the early Universe, by
the late-time accelerated expansion and nonsingular behavior in intermediate epochs, and classified them with respect to a number of transition points.
Transition points appear, when the acceleration parameter changes the sign, providing the natural partition of the Universe's history into
epochs of accelerated and decelerated expansion. The strategy and results of numerical calculations are advocated by the qualitative analysis of the instantaneous phase portraits
of the dynamic system associated with the key equation for the dark energy pressure evolution.
\end{abstract}


\maketitle

\section{Introduction}

The concepts of dark energy (DE) and dark matter (DM) \cite{DE1}-\cite{DM3}
considered as two manifestations of one unified dark fluid \cite{DF1}-\cite{DF5}
are basic elements of modern cosmology  and astrophysics.
Because the dark fluid is estimated to accumulate about $95\%$ of the Universe's energy, the coupling of DE with DM predestine the
main features of the Universe evolution.
In the first part of the work \cite{antigauss} we introduced the so-called
Archimedean-type force, which is linear in the four-gradient of the DE pressure and acts on the DM particles. This force
is a relativistic generalization of the classical Archimedean force and belongs to the class of effective forces described in \cite{BZ1}-\cite{BZ6}.
From mathematical point of view the presented model gives us a new self-consistent nonlinear scheme of interaction between two constituents of the
cosmic medium. From physical point of view the Archimedean-type force is an effective redistributor of the total energy of the
Universe between the DE and DM constituents.
The Archimedean-type model can be also considered in terms of two-fluid
representation of the cosmic medium \cite{M1}-\cite{M6}, in which the interaction terms $\pm Q$ appear in the
right-hand sides of separate balance equations for the DE and DM
with opposite signs and which disappear in a sum, when one deals with
the total balance equation.
We formulated in \cite{antigauss} the theory of interaction between DE and DM by
using the relativistic hydrodynamics for dark energy and
relativistic kinetics for dark matter.

In the first paper \cite{antigauss} we focused on the submodels with some special values of guiding parameters, which admit exact analytical solutions of the total
self-consistent system of master equations; in particular, we discussed the so-called anti-Gaussian solution. Now we
consider the results of numerical and qualitative analysis for all admissible sets of guiding parameters and initial data.
This paper is organized as follows.
In Sec.II we recall briefly the key equations and basic formulas of the model with Archimedean-type coupling between dark energy and dark matter.
In Sec.III the numerical results are presented: in Sec.III A we explain the scheme of numerical analysis and details of results representation;
in Sec.III B we classify the models using the number of transition points and discuss details of seven submodels, the perpetually accelerated
and periodic universes being among them. In Sec.IV we analyze the model of Archimedean-type coupling in terms of dynamic system associated with nonlinear key equation
for the DE pressure; in Sec.IV A we study the toy model, which relates to the autonomous dynamic system, find critical points and discuss two basic phase portraits
of this system; in Sec.IV B we analyze instantaneous phase portraits of a general nonautonomous dynamic system in the context of numerical results presented in Sec.III;
in Sec.IV C we consider asymptotical behavior of the models. Sec.V contains discussions: in Sec.V A we attract an attention to the inflationary behavior of the model
in the early Universe; in Sec.V B we show that all the models under discussion manifest the late-time accelerated expansion; in Sec.V.C we touch on the coincidence problem.

\section{Master equations of the toy model}

\subsection{Dynamic parameters of the Universe}

In order to describe the Universe's evolution from the
dynamic point of view, we find three key functions: the
scale factor $a(t)$, the Hubble function $H(t)$ and the
acceleration parameter ${-}q(t)$ given by the standard definitions,
\begin{equation}
H(t)\equiv \frac{\dot{a}(t)}{a(t)} \,, \quad -q(t)\equiv
\frac{\ddot{a}(t)}{a(t)H^2(t)} \,. \label{Hq}
\end{equation}
The space-time is considered to be of the spatially homogeneous
Friedmann-Lema$\hat{i}$tre-Robertson-Walker (FLRW) type with the metric
\begin{equation} ds^2 = dt^2 -a^2(t)[(dx^1)^2 + (dx^2)^2
+(dx^3)^2] \,. \label{metric}
\end{equation}
When all the functions describing the state of the Universe, ${\cal S}(t)$, depend on time through the
scale factor ${\cal S}[a(t)]$, it is convenient to use the new variable $x$ and the following
relations, associated with $x$:
\begin{equation}
x \equiv \frac{a(t)}{a(t_0)} , \  \frac{d}{dt} = x
H(x)\frac{d}{dx} , \ t{-}t_0 = \int_0^{\frac{a(t)}{a(t_0)}}
\frac{dx}{x H(x)} . \label{x}
\end{equation}
The last equality gives the scale factor $a(t)$, when $H(x)$ is
found from the Einstein equations, which can be written in the
form
\begin{equation}
x \frac{d}{dx} H^2(x) = - 8\pi G[\rho(x) {+} E(x) {+} \Pi(x) {+}
P(x)] \,,
\label{EinREDU1}
\end{equation}
\begin{equation}
H^2 = \frac{8\pi G}{3}(\rho {+} E) \,.
\label{EinREDU12}
\end{equation}
The quantities in the right-hand sides of the Einstein equations
are interpreted as follows: $\rho$ is the energy density of the
dark energy and $\Pi$ is its pressure; $E$ and $P$ are,
respectively, the total energy density and the total pressure of
the dark matter. In these terms the acceleration parameter $-q$
can also be represented as
\begin{equation}
{-}q(x)  {=} 1 {+} x \frac{d}{dx} \log{H(x)} {=} {-}
\frac{1}{2}\left[1{+}3 \left(\frac{\Pi{+}P}{\rho{+}E}\right)(x)
\right] . \label{q}
\end{equation}

\subsection{Balance equations}

It was shown in the first part of our work that the separate
equations of the energy balance for the DM and DE constituents of the dark fluid can be presented in the form
\begin{equation}
x \frac{d}{dx}E(x) + 3(E + P) = - {\cal Q}(x) \,, \label{bal1}
\end{equation}
\begin{equation}
x \frac{d}{dx}\rho(x) + 3(\rho + \Pi) =  {\cal Q}(x) \,.
\label{bal2}
\end{equation}
The total system (DE plus DM) is considered to be conserved, the sum of
(\ref{bal1}) and (\ref{bal2}) yields the total balance equation
\begin{equation}
x \frac{d}{dx} \left[\rho(x) {+} E(x) \right] + 3(\rho {+} E {+}
\Pi {+} P) =0 \,, \label{balance}
\end{equation}
which is, clearly, the compatibility condition for the pair of
equations (\ref{EinREDU1}) and (\ref{EinREDU12}). The source term ${\cal Q}$, which was
introduced in the first part of our work,  has the form
\begin{equation}
{\cal Q} \equiv 3 \ x \left[\frac{d}{dx}{\Pi}(x)\right]
\sum_{({\rm a})} {\cal V}_{({\rm a})} \ P_{({\rm a})}(x) \,,
\label{balance1alt2}
\end{equation}
where $P_{({\rm a})}$ is a partial pressure of the DM particles of
the sort $({\rm a})$ and ${\cal V}_{({\rm a})}$ is the guiding
parameter associated with the Archimedean-type force.

The DE energy density $\rho$ and DE pressure $\Pi$ are coupled by
the linear inhomogeneous equation of state \cite{EOS1}-\cite{REO1}
\begin{equation}
\rho(x) = \rho_0 + \sigma \Pi(x) + \xi x \frac{d}{dx} \Pi(x) \,.
\label{simplest0}
\end{equation}
The state functions $E(x)$ and $P(x)$, which characterize the DM
constituent, are presented by the integrals
\begin{equation}
E(x) = \sum_{({\rm a})}\frac{E_{({\rm a})}}{x^3} \int_0^{\infty}
q^2 dq \sqrt{1{+}q^2 F_{({\rm a})}(x)} \ e^{{-}\lambda_{({\rm a})}
\sqrt{1{+}q^2}}\,, \label{e(x)}
\end{equation}
\begin{equation}
P(x) = \sum_{({\rm a})}\frac{E_{({\rm a})}}{3x^3} \int_0^{\infty}
\frac{F_{({\rm a})}(x) q^4 dq}{\sqrt{1{+}q^2 F_{({\rm a})}(x)}} \
e^{{-}\lambda_{({\rm a})} \sqrt{1{+}q^2}}\,,  \label{p(x)}
\end{equation}
where the following auxiliary quantities are introduced
\begin{equation}
F_{({\rm a})}(x) = \frac{1}{x^2} \exp{\{2{\cal V}_{({\rm a})}
[\Pi(1){-}\Pi(x)]\}} \,, \label{FF}
\end{equation}
\begin{equation}
E_{({\rm a})} \equiv \frac{N_{({\rm a})} m_{({\rm a})}
\lambda_{({\rm a})}}{K_2(\lambda_{({\rm a})})}\,,  \quad
\lambda_{({\rm a})} \equiv \frac{m_{({\rm a})}}{k_{({\rm
B})}T_{({\rm a})}} \,, \label{E}
\end{equation}
\begin{equation}
K_{\nu}(\lambda_{({\rm a})}) \equiv \int_0^{\infty} dz \cosh{\nu
z} \cdot \exp{[-\lambda_{({\rm a})} \cosh z]}   \,. \label{McD}
\end{equation}
Here $K_2(\lambda_{({\rm a})})$ is the modified Bessel function,
$k_{({\rm B})}$ is the Boltzmann constant, and $T_{({\rm a})}$ is the
partial temperature of the particles of the sort $({\rm a})$ \cite{kin1,kin2}.

\subsection{Key equation of the toy model}

Clearly, the key element of the toy model is the function
$\Pi(x)$; to find the DE pressure $\Pi(x)$ we use the key equation
\begin{equation}
\xi x^2 \Pi^{\prime \prime}(x) {+} x \Pi^{\prime}(x) \left(4 \xi {+}
\sigma \right) {+} 3 (1{+}\sigma)\Pi {+}
3 \rho_0 {=} {\cal J}(x) ,
\label{key1}
\end{equation}
where the prime denotes the derivative with respect to $x$. The
source term ${\cal J}(x) ={\cal J}(x,\Pi{-}\Pi(1),\Pi^{\prime})$ is generally presented
by the integral
\begin{equation}
{\cal J}(x) {=} {-} \sum_{({\rm a})} E_{({\rm
a})} \frac{\left[x^2 F_{({\rm a})}(x)\right]^{\prime}}{2 x^4}
\int_0^{\infty}\frac{q^4 dq \ e^{{-}\lambda_{({\rm a})} \sqrt{1{+}q^2}}}{\sqrt{1{+}q^2 F_{({\rm a})}(x)}}
 , \label{key2}
\end{equation}
and we use them for numerical calculations. For the qualitative
analysis we use the explicit model expression for the source term
\begin{equation}
{\cal J}_{*}(x) = E_{*} {\cal V}_{*}
\frac{\Pi^{\prime}(x)}{x^{\mu}} \exp{\{{\cal V}_{*}
[\Pi(1){-}\Pi(x)]\}} \,. \label{J}
\end{equation}
Here $\Pi(1)$ denotes the initial value for the DE pressure (let us remind that $\Pi(x{=}1) = \Pi[a(t{=}t_0)]$).
When we deal with massless particles, the corresponding source
term coincides with (\ref{J}), if
\begin{equation}
\mu =3 \,, \quad  E_{*} = E_{(0)} \,, \quad {\cal V}_{*} = {\cal
V}_{(0)} \,. \label{key3}
\end{equation}
When we deal with the cold dark matter, we use the source term
(\ref{J}) with
\begin{equation}
\mu = 4 \,, \quad E_{*} = \frac{3 N_{({\rm C})}}{k_{({\rm
B})}T_{({\rm C})}} \,, \quad {\cal V}_{*} = 2 {\cal V}_{({\rm C})}
\,. \label{nrP00key}
\end{equation}
Thus, the key equation (\ref{key1}) is the ordinary differential
equation of the second order linear in the derivatives of the
unknown function $\Pi^{\prime \prime}(x)$, $\Pi^{\prime}(x)$ and
nonlinear in the function $\Pi(x)$ itself.

Let us mention that, when $\sigma \neq -1$, the effective guiding parameter
$\rho^{*}_0$ can be introduced instead of $\rho_0$. Indeed, using a new
unknown function $\pi(x) \equiv \Pi(x){-}\Pi(1)$ one can rewrite the key
equation (\ref{key1}) in the form
\begin{equation}
\xi x^2 \pi^{\prime \prime}(x) {+} x \pi^{\prime}(x) \left(4 \xi {+}
\sigma \right) {+} 3 (1{+}\sigma)\pi {+}
3 \rho^{*}_0 {=} {\cal J}(x,\pi,\pi^{\prime}),
\label{key1*}
\end{equation}
where
\begin{equation}
\rho^{*}_0 = \rho_0 + (1+\sigma)\Pi(1) \,.
\label{key1**}
\end{equation}
Clearly, the parameters $\rho^{*}_0$ and $\rho_0$ coincide at $\sigma {=}
{-}1$.

\section{Numerical results}

\subsection{About the scheme of analysis}

We analyze the Archimedean-type model as follows. First, we integrate the key
equation (\ref{key1}) with the source term (\ref{key2}) and find the DE pressure
$\Pi(x)$. Then we calculate the function $\rho(x)$ using
(\ref{simplest0}), the functions $E(x)$ and $P(x)$ using
(\ref{e(x)}) and (\ref{p(x)}), respectively, then we find the
Hubble function $H(x)$ using (\ref{EinREDU12}), the acceleration
parameter $-q(x)$ using (\ref{q}) and finally the scale factor
$a(t)$ using (\ref{x}). Let us make seven remarks about the scheme of numerical
analysis.

\noindent
1. The solution for the DE pressure $\Pi(x)$ depends on the following
guiding parameters: $\xi$, $\sigma$, $\rho^{*}_0$, ${\cal V}_{({\rm
a})}$, $E_{({\rm a})}$, $\lambda_{({\rm a})}$ and on two initial
values $\Pi(1)$, $\Pi^{\prime}(1)$. We fulfilled systematic numerical calculations
by varying one of these parameters and fixing the remaining ones, and then we classified
the obtained plots picking out principally different classes of solutions. For the numerical analysis we
used two-component DM model, one component being massless (ultrarelativistic), another one being cold
(nonrelativistic). Nevertheless, in order to illustrate the main details of
the Universe expansion on the plots we considered only one (leading) component of the DM. Below we
summarize the results, emphasize the main features, but do not discuss details of numerical calculations.

\noindent
2. For a number of sets of the  guiding parameters and initial data, mentioned above,
the behavior of the solution is of a singular type, i.e., the scale factor, Hubble function,
or/and DE pressure, DE energy, etc., can take infinite values at a finite time moment $t{=}t_{(s)}>t_0$; we do not
discuss such solutions here.

\noindent
3. We omit the discussion of submodels for which the DM energy density and DM pressure grow
infinitely at $t \to \infty$ or tend to finite (nonvanishing) asymptotic values. In other words, we
restrict ourselves by the submodels, for which the DE component dominates over DM starting from definite time moment; this motive explains
why we consider the parameters $E_{({\rm a})}(t_0)$ to be unchanged in the presented plots.

\noindent
4. There exist a lot of sets of the guiding parameters for which the solutions for the DE energy and DE pressure
are asymptotically unstable; we do not focus on them here.

\noindent
5. We discuss only the solutions for which the total energy $\rho{+}E$ is
non-negative for arbitrary $t$, thus guaranteeing that $H$ is a real function [see
(\ref{EinREDU12})].

\noindent
6. In order to describe the Archimedean-type model completely we have to
indicate the domain in the multidimensional space of parameters
$\xi$, $\sigma$, $\rho_0$, ${\cal V}_{({\rm
a})}$, $E_{({\rm a})}$, $\lambda_{({\rm a})}$, $\Pi(1)$  and $\Pi^{\prime}(1)$
in which the solutions are nonsingular and asymptotically stable; we hope to solve this problem in a special
paper and do not discuss them here.

\noindent
7. Figures 1-7 present the results of numerical calculations and are organized as follows. In the first
panels labeled (a) [$1 \leq N \leq 7$] the plot of the DE energy $\rho(\tau)$ is
presented; (b) contains the plot of the DM
energy density $E(\tau)$; the DE pressure $\Pi(\tau)$ and DM pressure $P(\tau)$ are presented in panels (c) and
(d), respectively; the plot of the Hubble function $H(\tau)$ can be found in panels
(e); the acceleration parameter $-q(\tau)$ is placed in (f), and
finally, the plots of the scale factor $a(t)$ are given in (g). Let
us remind, that the parameter $\tau \equiv \log{x} {=} \log{\frac{a(t)}{a(t_0)}}$
specifies the logarithmic scale in figure panels (a)-(f), and the value $\tau{=}0$ corresponds to $t{=}t_0$; the plots in panels (g) are
presented in terms of logarithm of the normalized cosmological time $t/t_0$.

\subsection{Multistage character of the Universe's evolution}

The analysis of the pictures presented in Figs.1(f),2(f),3(f),4(f),5(f),6(f), and 7(f) shows explicitly that
for a wide choice of sets of the
parameters $\xi$, $\sigma$, $\rho_0$, ${\cal V}_{({\rm a})}$, $E_{({\rm a})}$,
$\lambda_{({\rm a})}$, $\Pi(1)$ and $\Pi^{\prime}(1)$ the function
$-q(\tau)$ is not monotonic and can change the sign.
We indicate the points, in which the function $-q(\tau)$ changes the sign, as {\it transition} points, which divide the
history of the Universe into deceleration/acceleration {\it epochs}; we denote these points as
$\tau^{(k)}_{({\rm trans})}$.
The epochs with accelerated expansion can give way to the epochs with deceleration, the number
of changes being finite or infinite. When the (continuous) curve $-q(\tau)$ has
{\it extrema}, it is natural to divide the corresponding epochs into {\it eras}, the number of eras relates to the
number of maximums and minimums. Thus, depending on the number of transition points (related to the number of (simple) roots
of the equation $q(\tau)=0$), we divide below all the models into three classes and into a number of sub-classes.

\subsubsection{Perpetually accelerated universe (no transition points)}

The first class of models relates to the case when the function $-q(\tau)$ is
non-negative for $\tau \geq 0$. In \cite{antigauss} we discussed the so-called anti-Gaussian model: exact
solutions in this model were presented in terms of elementary functions at $\sigma {=}{-}1$. In
particular, the  acceleration parameter obtained there,
${-} q(t) {=} 1 {+} \frac{3 {\cal V}_{(0)}}{ 16 \pi G (t{-}t^{*})^2}$,
is a monotonic positively defined function exceeding the value ${-}q_{({\rm dS})}{=}1$, the acceleration parameter
for the de Sitter models. In other words, there are no transition points, there is no splitting of the history into epochs
and eras, when we deal with an anti-Gaussian universe. Another example presented in \cite{antigauss} describes superexponential
expansion with the acceleration parameter given by
${-}q(t) {=} 1 {+} \sqrt{\frac{9\rho_0}{2\rho(1)}}  \frac{\sinh{\left[\sqrt{12\pi G \rho_0} \ (t{-}t_0) \right]}}{\cosh^2{\left[\sqrt{12\pi G \rho_0} (t{-}t_0) \right]}}$.
The plot of this function starts with ${-}q(t_0){=}1$, reaches the maximum ${-}q_{({\rm max})}{=}1 {+} \sqrt{\frac{9\rho_0}{8\rho(1)}}$ at $t{=}t_0 {+} \frac{\log{(1+\sqrt2)}}{\sqrt{12\pi G \rho_0}}$, and tends asymptotically to ${-}q(\infty){=}1$. In the context of our terminology, we also deal with perpetually accelerated universe, but its history is now divided into two eras.
Here we discuss an alternative example of a perpetually accelerated
universe with $\sigma \neq {-1}$, studied numerically.
Typical pictures illustrating this class of models are given in Fig.1.

\begin{figure}
\centerline{\includegraphics{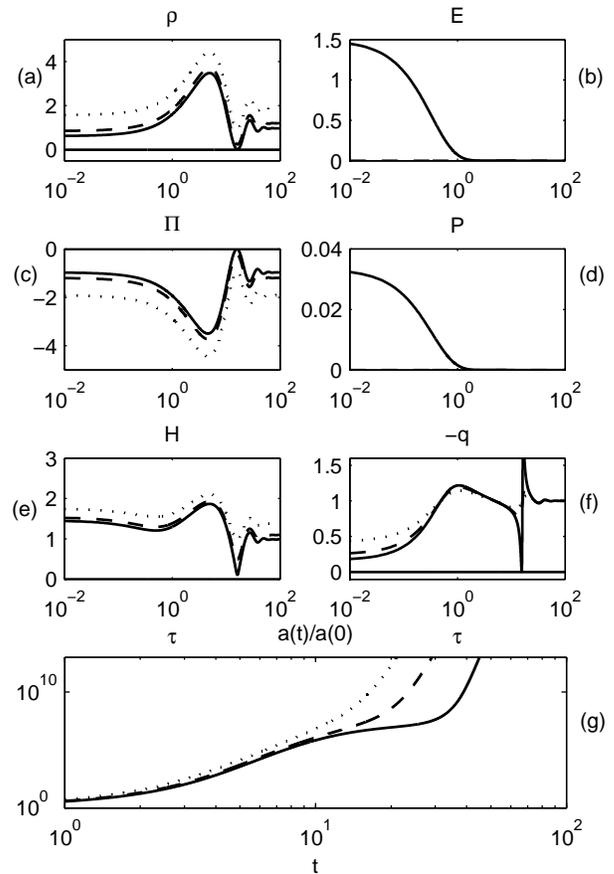}}
\caption {{\small Perpetually accelerated universe: the acceleration parameter
${-}q(\tau)$ is non-negative for arbitrary $\tau$. Since there are no transition points, in which ${-}q(\tau)$ changes
the sign, the history of this universe
includes only one epoch; since at least five extrema can be recognized in Fig.1(f), the history consists of six or more
eras. Here the dark solid line relates to the following set of the parameters:
$\xi{=}0.35$, $\sigma {=}{-}0.99$ (i.e., $3\xi{+}\sigma \simeq 0.06>0$),
${\cal V}_{(0)} {=}1$, $E_{(0)} {=} 0.0205$, $\lambda_{(0)} {=}1$ (i.e., the DM is initially medium-relativistic),
$\rho_{*} {=}0.333 \cdot 10^{-4}$, and $\Pi^{\prime}(1){=}{-}1$.}}
\end{figure}
The solid curve in Fig.1(f) touches the line $q=0$ in one point, other curves lie above
$q=0$, thus there are no transition points. In this class of models the acceleration parameter $-q$ as a function of
time is non-negative for arbitrary $\tau$: this means that the accelerated expansion of the Universe is perpetual.
For the model under discussion the DE energy density $\rho$ is (perpetually) non-negative [see Fig.1(a)],
the DE pressure $\Pi$ is nonpositive.
The acceleration parameter, the DE energy density and pressure, and the Hubble function
change nonmonotonically with time, thus, the statefinder
parameter $r \equiv \frac{1}{a H^3} \frac{d^3 a}{dt^3}$, introduced in
\cite{statefinder}, is necessary to describe the details of the Universe's
evolution. We do not focus now on these fine details of the model and their physical interpretation,
nevertheless, we have to stress, that extrema of the function
$-q(\tau)$ divide naturally the history of the Universe into the
corresponding eras, thus describing the multistage character of its evolution.

The history of a perpetually accelerated universe includes only one epoch, but consists at
least of six eras. Indeed, one can clearly identify at least three maximums and two minimums of the curve $-q(\tau)$ which divide the
acceleration epoch into six eras.
The first era ($\tau <10^{0}$) can be indicated as the era of a first superacceleration. During this era
the DE energy - density, $\rho(\tau)$, grows monotonically [see Fig.1(a)], the DM energy density, $E(\tau)$,
decreases [see Fig.1(b)], but at the beginning of the era DM dominates over the DE from the point of view of
energy contribution.
The second era finishes with slumps of the DE
energy density, of the Hubble function and of the acceleration parameter. The DE pressure, which reached its global
maximum at this moment, becomes, clearly, a locomotive power of further universe evolution.
The third era can also be indicated as the era of super-acceleration, which is characterized by the growth of DE
energy density and the Hubble parameter. Fourth, fifth, etc. eras can be
considered in terms of relaxation to the state with asymptotically constant
positive values $-q_{\infty}$, $H_{\infty}$, $\rho_{\infty}$ and negative
$\Pi_{\infty}$. The behavior of the DM state functions $E(\tau)$ and $P(\tau)$
during the second, third, etc. eras  can not be recognized in the plots in
Fig.1(b) and Fig.1(d), that is why we visualize their behavior specially in the discussion
below using more appropriate scale (see Fig.11 and Fig.12).

\subsubsection{Periodic  models (infinite number of transition points)}

We indicate the model by the term periodic when the equation $q(\tau)=0$ has an infinite number of
roots and the history of the Universe splits into infinite number of identical epochs with accelerated and decelerated
expansion. Clearly, it is the case opposite to the first one (perpetually accelerated universe).
Figure 2 illustrates this terminology:  the DE energy density
and DE pressure, as well as the Hubble function and acceleration
parameter oscillate with fixed frequency and amplitude (we should keep in mind that the time scale
in Fig.2 (a)-(f) is logarithmic, so that the frequency seems to be visually increasing).
In the model presented in Fig.2 the
DE energy density $\rho$ is non-negative, the averaged over time (mean) value of the DE pressure $<\Pi>$
is negative, and the sum $\rho+\Pi$ is positive.
The plots in the Fig.2(e) and Fig.(2)(f) demonstrate that the stable oscillatory regime
starts in the Universe after the second transition point. The first epoch
is characterized by the decelerated expansion and consists of three eras; at the end of the first era
($\tau \simeq 10^{0}$) the DM contribution into the total energy density becomes negligible in comparison with the DE
contribution. The second epoch relates to the accelerated expansion and can be split into two
eras, the maximal value of the acceleration parameter being about 3
times smaller than for the subsequent epochs.
The scale factor $a(t)$ oscillates near the de Sitter
curves [see Fig.2(g)]. The periodic model can be obtained as a limiting
case, when the parameter $3\xi {+} \sigma$ tends to its critical value $3\xi {+} \sigma \to 0_{+0}$.
\begin{figure}
\centerline{\includegraphics{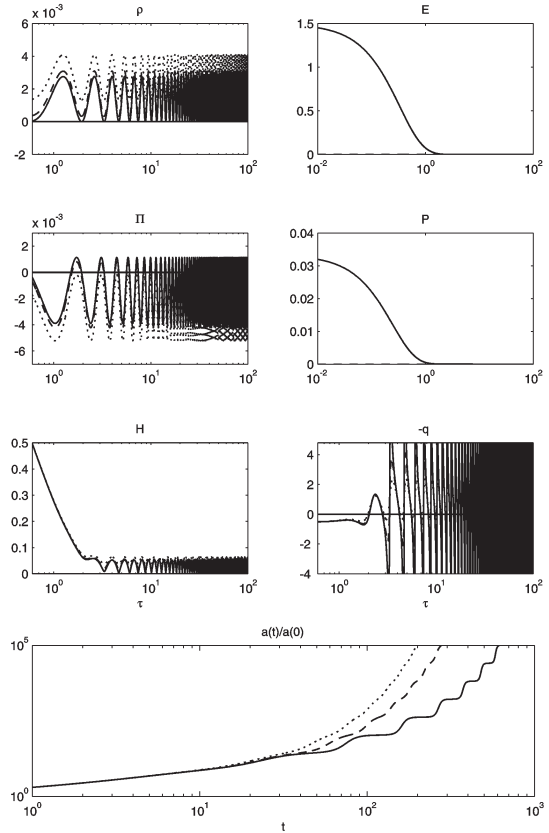}}
\caption {{\small Periodic universe with infinite number of acceleration/deceleration epochs. This model is
a limiting case, when the parameter $3\xi {+} \sigma$ tends to its critical value $0_{+0}$.
For this model not only the sum $\rho{+}E$, but the DE energy density
$\rho$ itself remain non-negative. Oscillations in
the plots of the Hubble function $H(\tau)$ and of the scale factor $a(t)$
are the most explicit (see dark solid line), when the parameters of the model are the following:
$\xi{=}0.1$, $\sigma {=}{-}0.299999$ (i.e., $3\xi{+}\sigma \simeq 10^{-6}>0$),
${\cal V}_{(0)} {=}1$, $E_{(0)} {=} 0.0205$, $\lambda_{(0)} {=}1$,
$\rho_{*} {=}0.333 \cdot 10^{-4}$, and $\Pi^{\prime}(1){=} 0.01$.
}}
\end{figure}

\subsubsection{Universe evolution with a finite number of transition points}

Let us focus on the models for which the equation $q(\tau)=0$ has a finite number of simple
roots, $m \geq 1$. This means that during the Universe evolution the epochs with  accelerated expansion
are changed $m$ times by the deceleration epochs, and vice versa. When the roots are not simple, special
attention to such models is necessary, since the existence of the root does not guarantee
the change of the sign of the function ${-}q(\tau)$ (e.g.,  when the root is double, the curve ${-}q(\tau)$ touches  the line $q=0$, and this root does not give a transition point).
The class of models with $m$ transition points can be naturally
divided into subclasses with $m{=}1$, $m{=}2$, etc.; every subclass can be divided into sub-subclasses according
to physical and geometrical motivation. Let us focus on the basic ones.

\vspace{3mm}
\noindent
{\it (i) One transition point.}

Figure 3 illustrates the subclass of models, for which the plot of the acceleration parameter $-q(\tau)$ (see Fig.3(f))
can be visualized as a deformed Heaviside step-function. The change of the
deceleration epoch by the acceleration epoch takes place only once in a narrow time
period, and the plot looks like a typical picture for a phase transition.
Taking into account the plot of the Hubble function [see Fig.3(e)] one can conclude that the Universe's evolution starts
and finishes on quasi de Sitter asymptotes with different Hubble constants $H_0 \neq H_{\infty}$.
The functions $\rho(\tau)$, $\Pi(\tau)$, related to the DE component of the dark fluid, monotonically increase, and the
functions $E(\tau)$, $P(\tau)$, related to the DM, monotonically decrease, the sum $\rho(\tau){+}E(\tau)$ being non-negative
for arbitrary $\tau >0$. The second epoch, characterized by
the accelerated expansion,  looks like the de Sitter-type stage with $\rho+\Pi=0$ and vanishing $E$ and
$P$. The first and second epochs are not divided into eras in the framework
of this model.

\begin{figure}
\includegraphics{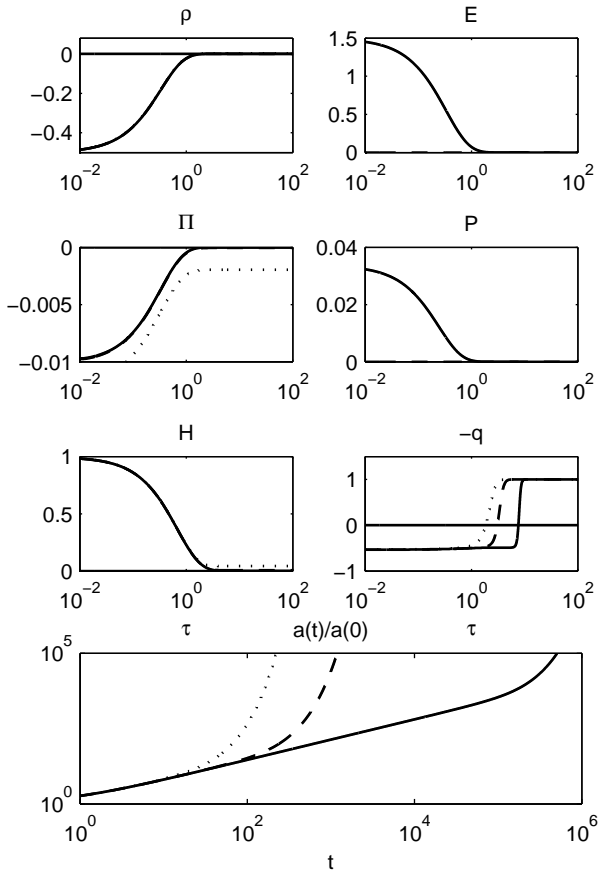}
\caption {{\small The model with one transition point and ${-}q(\tau)$ curve of the Heaviside step-function type.
The first (deceleration) epoch and the second (acceleration) epoch are not
divided into eras. The epoch
with the accelerated expansion is of the de Sitter-type, characterized by $\rho{+}\Pi{=}0$. The sum $\rho(\tau){+} E(\tau)$
is non-negative for arbitrary $\tau >0$. The parameters of the model are the following:
$\xi{=}0.1$, $\sigma {=}50$,
${\cal V}_{(0)} {=}1$, $E_{(0)} {=} 0.0205$, $\lambda_{(0)} {=}1$,
$\rho_{*} {=}0.333 \cdot 10^{-4}$, and $\Pi^{\prime}(1){=} {-} 5$.}}
\end{figure}

Figure 4 illustrates an alternative subclass of models with one transition point.
In contrast to the first example, the functions $\rho(\tau)$, $\Pi(\tau)$, $H(\tau)$, $P(\tau)$ and ${-}q(\tau)$
have extrema, so that one can pick out two eras in both epochs of the Universe history. This model also describes the
transition from one de Sitter-type universe to another one with different
Hubble constants $H_0 \neq H_{\infty}$. It is important to stress that the
DM pressure $P(\tau)$ is not monotonic now: in contrast to the DM energy
density, the DM pressure grows in the early Universe, reaches the maximum and then decreases quickly
in the deceleration epoch.

\begin{figure}
\centerline{\includegraphics{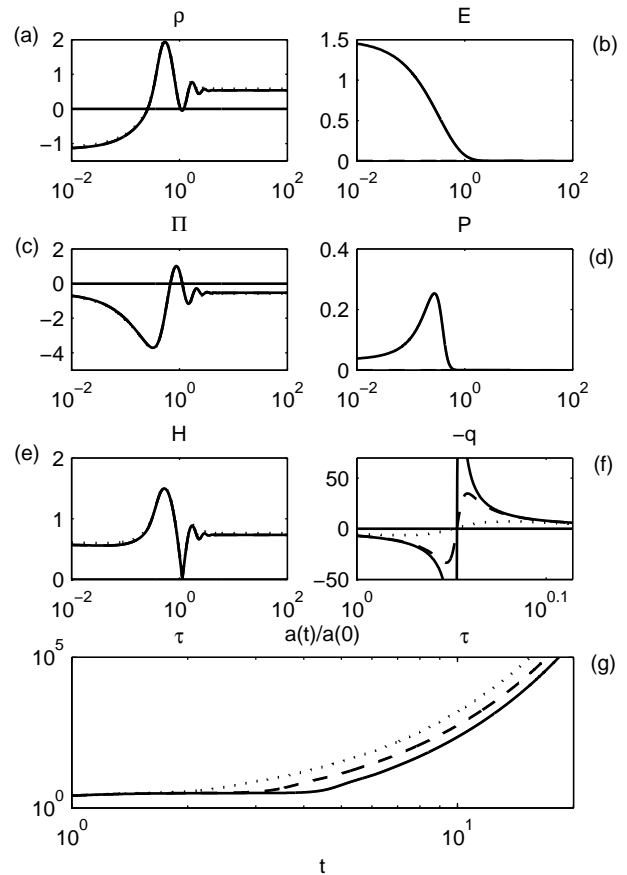}}
\caption {{\small The model with one transition point and two eras in both decelerated and accelerated epochs.
The model describes the transition from one de Sitter-type universe to another one with different
Hubble constants $H_0 \neq H_{\infty}$. In contrast to DM energy
density, the DM pressure grows in the early universe, reaches the maximum and then decreases quickly
in the deceleration epoch.
The parameters of the model are the following:
$\xi{=}0.1$, $\sigma {=}0$,
${\cal V}_{(0)} {=}1$, $E_{(0)} {=} 0.0205$, $\lambda_{(0)} {=}1$,
$\rho_{*} {=}0.333 \cdot 10^{-4}$, and $\Pi^{\prime}(1){=} {-}17$.
}}
\end{figure}

\vspace{3mm}
\noindent
{\it (ii) Two transition points.}

The example of the model with two transition points is given in Fig.5.
The behavior of the functions $\rho(\tau)$, $\Pi(\tau)$, $E(\tau)$, and $H(\tau)$
is similar to the one, given in Fig.1, but there are new details in the panels (d) and (f).
First, the DM pressure reaches maximum at the end of the first era of the first acceleration
epoch. The second new detail is that the deceleration epoch appears, which separates two acceleration epochs with
different asymptotic values of the Hubble function.

\begin{figure}
\centerline{\includegraphics{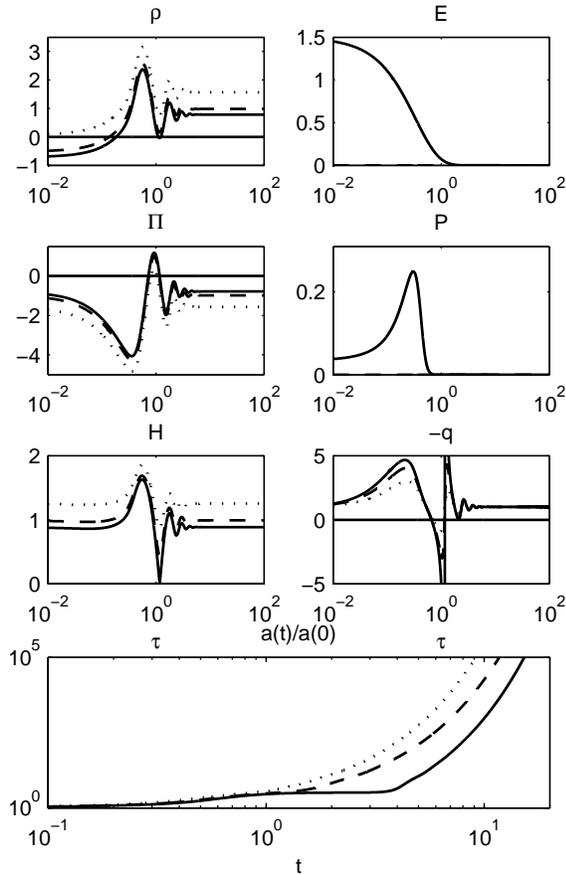}}
\caption {{\small The model with two transition points, two acceleration epochs and one deceleration epoch.
The plots for the DE energy density, DE pressure, and Hubble functions are analogous to the curves given in Fig.1.
The DM pressure has a maximum at the end of the first era of the first acceleration
epoch in analogy with the curve given by Fig.4(d).
The parameters of the model are the following:
$\xi{=}0.1$, $\sigma {=}{-}0.08$,
${\cal V}_{(0)} {=}1$, $E_{(0)} {=} 0.0205$, $\lambda_{(0)} {=}1$,
$\rho_{*} {=}0.333 \cdot 10^{-4}$, and $\Pi^{\prime}(1){=} {-}15$.
}}
\end{figure}

\vspace{3mm}
\noindent
{\it (iii) Three transition points.}

The model which displays two epochs of decelerated expansion and two epochs
of accelerated evolution, is presented by Fig.6. The
graph of the function ${-}q(\tau)$ is of the so-called ${\cal N}$  type.
The start of the Universe's expansion relates to the deceleration epoch, which is then replaced by a short
acceleration epoch as in the case of the periodic model [compare Fig.6(f) with
Fig.2(f)]. The Universe's evolution during the second epoch of decelerated expansion and the subsequent second
accelerated epoch is analogous to the model with one transition point (compare Fig.6(f) with
Fig.3(f)). Clearly, the second (final) accelerated epoch is of the de Sitter type.

\begin{figure}
\centerline{\includegraphics{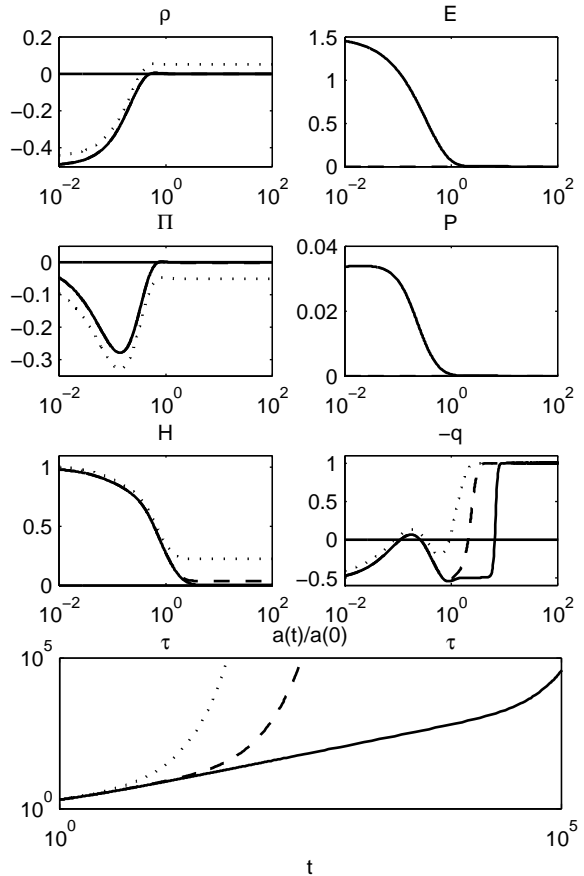}}
\caption {{\small The model with two epochs of decelerated expansion and two epochs
of accelerated evolution. The plot of the function ${-}q(\tau)$ is of the so-called ${\cal N}$  type.
The second accelerated epoch is of the de Sitter type.
The parameters of the model are the following:
$\xi{=}0.1$, $\sigma {=}1$,
${\cal V}_{(0)} {=}1$, $E_{(0)} {=} 0.0205$, $\lambda_{(0)} {=}1$,
$\rho_{*} {=}0.333 \cdot 10^{-4}$, and $\Pi^{\prime}(1){=} {-}5$.
}}
\end{figure}

\vspace{3mm}
\noindent
{\it (iv) Quasiperiodic evolutionary models.}

Figure 7 illustrates the behavior of the model with a big but finite number
of transition points. The first, second, and third epochs of deceleration, as well as the first, second, and third epochs of
accelerated expansion in this model have the same character as for the periodic model [compare with Fig.2(f)]. Nevertheless,
starting from some transition point with the number $(k)$, i.e.,
when $t>t^{(k)}_{({\rm trans})}$, the curve ${-}q(\tau)$ remains above the line $q{=}0$,
thus, later the Universe's expansion is accelerated. Clearly, this model is intermediate
between the model with two transition points [see Fig.5] and periodic model [see
Fig.2]. The behavior of the functions $\rho(\tau)$,  $\Pi(\tau)$, $H(\tau)$
and ${-}q(\tau)$ is quasiperiodic: the amplitudes of their oscillations decrease asymptotically.
Late-time behavior in average is of the de Sitter type with positive DE
energy density and negative DE pressure. The interest in such models has been renewed after publication of the paper \cite{Penrose}.

\begin{figure}
\centerline{\includegraphics{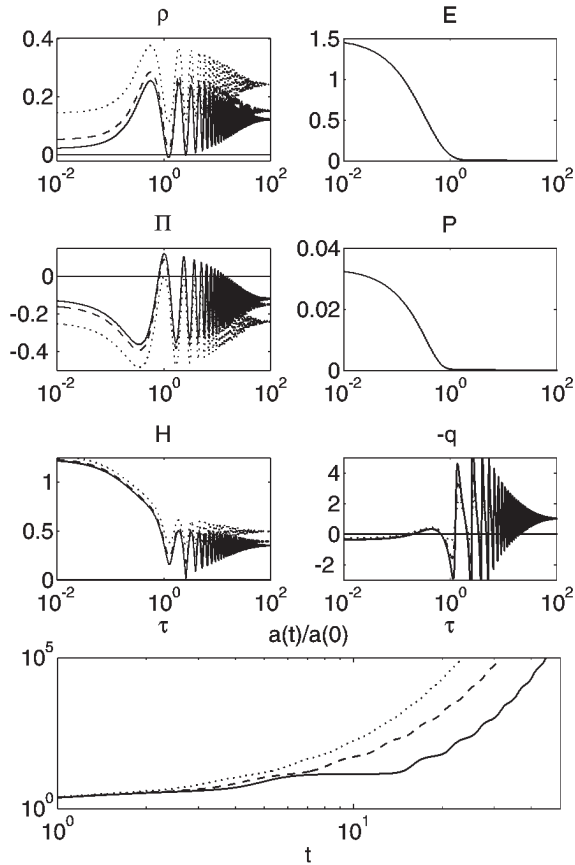}}
\caption {{\small The intermediate model with big but finite number of transition points. The oscillatory regime
is produced by the DE component of the dark fluid, and it is
switched on at the moment when the DM energy density and DM pressure become vanishing. Late-time expansion is, evidently,
accelerated. The dark solid lines relate to the following parameters:
$\xi{=}0.1$, $\sigma {=}{-}0.29$ (i.e., $3\xi{+}\sigma \simeq 0.01>0$),
${\cal V}_{(0)} {=}1$, $E_{(0)} {=} 0.0205$, $\lambda_{(0)} {=}1$,
$\rho_{*} {=}0.333 \cdot 10^{-4}$, and $\Pi^{\prime}(1){=} {-}1$.
}}
\end{figure}

\section{Qualitative analysis}

In Sec.III we classified the solutions obtained numerically
according to the number of transition points, in which the decelerated expansion of the Universe is changed by the
accelerated expansion and vice versa. In order to explain the results qualitatively, below  we
analyze the key equations by two methods: constructing the phase portraits and using asymptotic analysis.
First of all, we consider the example of the model with $\sigma={-}1$, for which the key equation
can be reduced to the {\it autonomous} dynamic system; we find critical points and draw the corresponding  phase
portraits. On the basis of these results we study nonautonomous dynamic systems with
$\sigma \neq{-}1$ taking into account that critical points of this nonautonomous system
remain of the same type as for the case $\sigma {=} {-}1$, but they drift with time on
the phase plane. Finally, we analyze three asymptotic limits: $\Pi \to \infty$, $\Pi \to 0$ and $\Pi \to const \neq 0$,
which allow us to simplify the key equation significantly.

\subsection{Dynamic system associated with the key equation}

Let us introduce new variables $X(\tau)$, $Y(\tau)$ and $\tau$ as follows
$$
X = \frac{E_{*}{\cal V}_{*}}{\xi (\mu {+}1)} x^{-\mu -1} \exp\{-{\cal V}_{*} [\Pi(x)-\Pi(1)] \} \,,
$$
\begin{equation}
Y \equiv \frac{{\cal V}_{*}}{(\mu{+}1)} x \ \frac{d \Pi}{dx} \,, \quad \tau \equiv (\mu {+}1)\log{x} \,. \label{Q2}
\end{equation}
Then the key equation of the second order (\ref{key1}) with the source term (\ref{J}) can be rewritten as a
pair of dynamic equations of the first order:
$$
\frac{dX}{d\tau} = - X \left(1 + Y \right) \,,
$$
\begin{equation}
\frac{dY}{d\tau} {=} {-} A {+} Y (X{-}B) {+} \frac{3(1{+}\sigma)}{\xi(\mu{+}1)^2}\left[
\tau {+} \log{\left( \frac{\xi (\mu{+}1)}{E_{*}{\cal V}_{*}} X \right)}
\right]
, \label{Dq1}
\end{equation}
where
\begin{equation}
A \equiv \frac{3\rho^{*}_0 {\cal V}_{*}}{\xi (\mu + 1)^2}\,, \quad B
\equiv \frac{3\xi+\sigma}{\xi (\mu{+}1)} \,. \label{Q6}
\end{equation}
The quantity $X(\tau)$ is non-negative due to its definition (\ref{Q2}).
$X$ tends to zero in two different cases: first, when $\Pi \to \infty$,
or second, asymptotically at $x \to \infty$, if $\Pi(\infty)$ is finite.
The initial value for the parameter $\tau$,
$\tau{=}0$ relates to the moment $t{=}t_0$ and, respectively, $x{=}1$,
then initial data for the integral curves in terms of these variables are the following:
\begin{equation}
X(0)= \frac{E_{*}{\cal V}_{*}}{\xi (\mu {+}1)} \,, \quad Y(0)= \frac{{\cal V}_{*}}{(\mu{+}1)}\Pi^{\prime}(1)
\,. \label{Q3}
\end{equation}
Generally, this dynamic system is nonautonomous because of the term linear in $\tau$ in the second equation.
Nevertheless, there is a case $\sigma {=} {-}1$ when the system (\ref{Dq1}) becomes
autonomous. Let us consider this case in more detail.

\subsubsection{Autonomous dynamic system at $\sigma={-}1$}

The autonomous dynamic system
\begin{equation}
\frac{d X}{d \tau} = - X \left[1 +
Y\right] \,, \quad \frac{d Y}{d \tau} = - A
+ Y \left[X - B \right]  \label{Q5}
\end{equation}
contains only two control parameters: $A$ and $B$, the parameter $A$ being now positive.
The new dimensionless parameter
\begin{equation}
\theta \equiv \frac{B}{A} = \frac{(3\xi-1)(\mu + 1)}{3\rho_0 {\cal V}_{*}}  \label{Q61}
\end{equation}
plays an interesting role in further analysis.
Physically appropriate integral curves should be situated in the domain $X \geq 0$ on the plane $X0Y$.
We consider the integral curves, which penetrate into the domain $X<0$,
as nonadmissible and omit the models with the corresponding set of guiding parameters.
The qualitative analysis of the two-parameter autonomous dynamic
system (\ref{Q5}) demonstrates the following interesting features.

\subsubsection{Critical points}

When $A \neq B$ and $B\neq 0$, there are two critical points associated with this dynamic
system. The first critical point
\begin{equation}
X_{(1)} = 0 \,, \quad Y_{(1)} = - \frac{A}{B} = - \frac{1}{\theta}
\label{D1}
\end{equation}
is the saddle one, when $\theta<1$,
and is the stable nodal point, when $\theta>1$. The second critical point is
\begin{equation}
X_{(2)} = B-A = (\theta - 1) \frac{3\rho_0 {\cal V}_{*}}{\xi (\mu + 1)^2} \,,
\quad Y_{(2)} = -1 \,,
\label{D3}
\end{equation}
and it coincides with the first one, when $B{=}A$, i.e., when $\theta{=}1$. Generally, the second critical
point can be

\noindent
{\it (i)} the saddle point, when $\theta>1$ (the first critical point is
the nodal one);

\noindent
{\it (ii)} the stable nodal point, when $\frac{3\rho_0 {\cal V}_{*}}{4\xi (\mu {+} 1)^2}>1{-}\theta>0$
(the first critical point is
the saddle one);

\noindent
{\it (iii)} the stable focus, when $0<\frac{3\rho_0 {\cal V}_{*}}{4\xi (\mu {+} 1)^2}<1{-}\theta$
(the first critical point is
the saddle one); or

\noindent
{\it (iv)} the degenerated node, when $0<\frac{3\rho_0 {\cal V}_{*}}{4\xi (\mu {+} 1)^2}{=}1{-}\theta$
(the first critical point is the saddle one).

\subsubsection{Two phase portraits of the autonomous dynamic system}

A typical phase portrait of the autonomous dynamic system with a stable node
and a saddle point is presented in Fig.8. Integral curves in the northern and western
domains, which are cut out by the separatrices of the saddle point with coordinates $X_{(2)}{=}0.2125$, $Y_{(2)}{=}{-}1$, tend
asymptotically to the nodal point with $X_{(1)}{=}0$, $Y_{(1)}{=}{-}0.15$.
All the integral curves in the southern and eastern domains of this saddle
point can be asymptotically characterized by infinite values of $X$ and/or
$Y$.

\begin{figure}
\centerline{\includegraphics{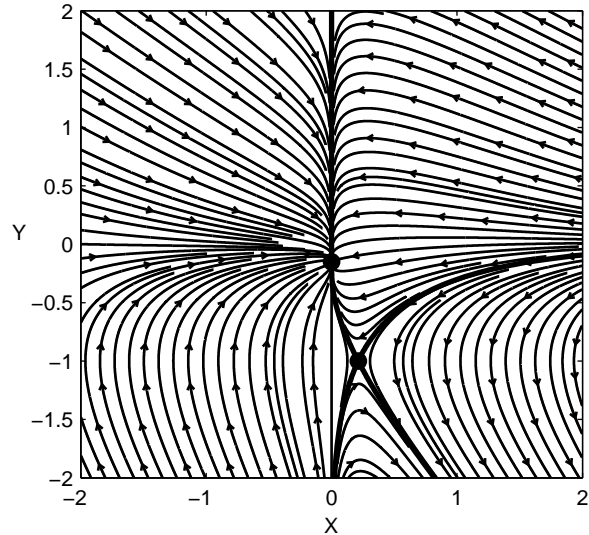}}
\caption {{\small
Phase portrait of the autonomous dynamic system with stable  node at $X_{(1)}{=}0$, $Y_{(1)}{=}{-}0.15$
and saddle point at $X_{(2)}{=}0.2125$, $Y_{(2)}{=}{-}1$. Integral curves in the northern and western
domains, which are cut out by the separatrices of the saddle point, tend
asymptotically to the node. Integral curves in the southern and eastern domains of this saddle
point can be asymptotically characterized by infinite values of $X$ and/or $Y$.
}}
\end{figure}

A typical phase portrait of the autonomous dynamic system with a focus and
a saddle point is given in Fig.9. The domain of quasiperiodic motion is
situated in the south-western domain, which is cut out by the separatrices of the saddle point
at $X_{(1)}{=}0$, $Y_{(1)}{=}0.25$. This quasiperiodic motion is associated with the presence of the
focus at  $X_{(2)}{=} -0.59375$, $Y_{(2)}{=}{-}1$. Curves walking in three other
domains are asymptotically unstable.

\begin{figure}
\centerline{\includegraphics{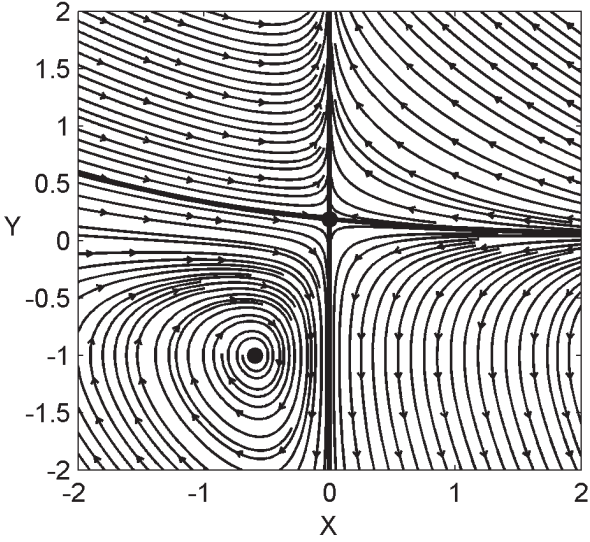}}
\caption {{\small
Phase portrait of the autonomous dynamic system with stable focus and
saddle point. The domain of quasiperiodic motion is
situated in the south western domain, which is cut out by the separatrices of the saddle point
at $X_{(1)}{=}0$, $Y_{(1)}{=}0.1875$. The focus is situated at $X_{(2)}{=} -0.59375$, $Y_{(2)}{=}{-}1$.
}}
\end{figure}

\subsection{On the behavior of the nonautonomous dynamic system with $\sigma>-1$}

When $\sigma{=}{-}1$ and the dynamic system is autonomous, there are two typical phase portraits, given by Fig.8 and Fig.9. The qualitative difference between them
is predetermined by specific combinations of the guiding parameters $\xi$, $\rho_0$, and ${\cal V}_{*}$: any changes on the phase space are associated with variations of these guiding parameters.
Typical phase portraits  with fixed guiding parameters remain unchanged with time, and the integral curves can be prolonged till $\tau \to \infty$.
When $\sigma>{-}1$, the dynamic system is nonautonomous, nevertheless, one can describe such a model qualitatively in terms of instantaneous phase portrait distortion,
in terms of (quasi)critical points drift, and their transformations. One can stress that in this case a restructuring of the instantaneous phase portraits can take place
because of time growth, when the guiding parameters remain unchanged.

\subsubsection{Critical points drift and transformation}

Let us consider a close vicinity of the point $\tau {=} \tau_0$ and put the value $\tau_0$ to
the right-hand side of the second equation (\ref{Dq1}). In the small vicinity of $\tau_0$
($\tau {=} \tau_0 {+} \eta$, and $\eta  << \tau_0 $ is a small deviation) we
can consider (quasi) critical points in analogy with true critical points, studied in the previous case. The
first (quasi) critical point relates to $X_{(1)}{=}0$, $Y_{(1)} {=} \infty$, when
$3\xi{+}\sigma<0$, and $X_{(1)}{=}0$, $Y_{(1)} {=} {-} \infty$, when
$3\xi{+}\sigma>0$. In both cases these points disappear from the admissible part of the dynamic
plane $X0Y$. Other (quasi) critical points relate to the value $Y_{(2)}{=}
{-}1$, and the corresponding values $X$ can be found from the transcendent equation
\begin{equation}
X - \beta \log{X} = {\cal K}(\tau_0) \,, \quad \beta \equiv
\frac{3(1{+}\sigma)}{\xi(\mu{+}1)^2} \geq 0\,,
\label{trans1}
\end{equation}
\begin{equation}
{\cal K}(\tau_0) {=}  \beta \left[\tau_0
{+}\log{\frac{\xi(\mu{+}1)}{E_{*}{\cal V}_{*}}} {+}
\frac{(3\xi{+}\sigma)(\mu{+}1){-}3 \rho^{*}_{0}{\cal V}_{*}}{3(1{+}\sigma)}\right].
\label{trans2}
\end{equation}
The plot of the function $f(X) \equiv X {-} \beta \log{X}$ is presented in
Fig.10.
\begin{figure}
\centerline{\includegraphics{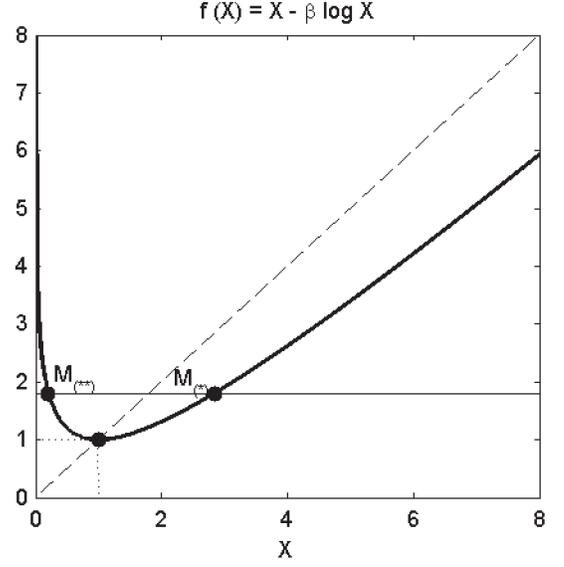}}
\caption {{\small The plot illustrates the appearance of a pair of (quasi)critical points, when the cosmological time increases.
The function $f(X) \equiv X {-} \beta \log{X}$ has the minimum
$f_{({\rm min})} {=} \beta\log{\frac{e}{\beta}}$ at $X{=}\beta$; the point of minimum divides the
plot of this function into two branches: first of them is situated at
$0<X<\beta$, the second one relates to the interval $X>\beta$.
Clearly, Eq. (\ref{trans1}) has no solutions, when ${\cal K}(\tau_0) < \beta \log{\frac{e}{\beta}}$; it has only
one (double) root, when ${\cal K}(\tau_0) {=} \beta \log{\frac{e}{\beta}}$; there are two different roots, when ${\cal K}(\tau_0) >
\beta \log{\frac{e}{\beta}}$, the corresponding points indicated as $M_{(*)}$ and $M_{(**)}$. The curve on the plot corresponds to the value $\beta=1$. }}
\end{figure}
The function $f(X)$ has the minimum
$f_{({\rm min})} {=} \beta\log{\frac{e}{\beta}}$ at $X{=}\beta$, which divides the
corresponding curve into two branches: the first of them is situated at
$0<X<\beta$ (left branch), and the second one relates to the interval $X>\beta$ (right branch).
Clearly, Eq. (\ref{trans1}) has no solutions, when
${\cal K}(\tau_0) < \beta \log{\frac{e}{\beta}}$ or equivalently when $\tau_0 < \tau_{({\rm crit})}$, where the critical time moment is
\begin{equation}
\tau_{({\rm crit})} {=} 1 {+} \frac{\rho_0^{*}{\cal V}_{*}}{1{+}\sigma} {-} \frac{(\mu{+}1)(3\xi{+}\sigma)}{3(1{+}\sigma)} {+} \log{\frac{(\mu{+}1){\cal V}_{*}E_{*}}{3(1{+}\sigma)}}
.
\label{trans309}
\end{equation}
Eq. (\ref{trans1}) has only one (double) root, when ${\cal K}(\tau_0) {=} \beta \log{\frac{e}{\beta}}$, or $\tau_0 {=} \tau_{({\rm crit})}$; there are two
different roots, when ${\cal K}(\tau_0) > \beta \log{\frac{e}{\beta}}$ or equivalently $\tau_0 > \tau_{({\rm crit})}$.
  In other words,  there are no (quasi)critical points at $\tau_0 < \tau_{({\rm crit})}$, one point appears at $\tau_0 {=} \tau_{({\rm crit})}$, and two (quasi)critical points
appear in the instantaneous phase portrait, when $\tau_0 > \tau_{({\rm crit})}$.

The first (quasi)critical point indicated as $M_{(*)}(\tau_0)$ drifts to infinity, since the value $X_{(*)}$, for which the right
branch of the curve happens to be crossed by the horizontal straight line, increases with $\tau_0$. In order to recognize the type of this (quasi) critical point,
let us find the roots of the characteristic equation associated with the dynamic system in the vicinity of the point with coordinates $(X_{(*)}, {-}1)$,
\begin{equation}
\lambda_{1,2} = \frac{1}{2}\left[X_{(*)}{-}B \pm \sqrt{(X_{(*)}{-}B)^2 {+} 4(X_{(*)}{-}\beta)} \right]
\,.
\label{trans3}
\end{equation}
At the critical moment of time the equality $X_{(*)}{=}\beta$ takes place, and we deal with the degenerated node; when $X_{(*)}> \beta$, there are two real roots of
different signs, thus the (quasi)critical point is the saddle one.

The second (quasi)critical point $M_{(**)}$ drifts to the point $(0,{-}1)$ when $\tau_0$ increases, since
the value $X_{(**)}\leq \beta$, for which the left branch of the curve crosses the horizontal straight line, tends to zero.
The type of the second (quasi)critical point depends on the value $\tau_0$ and on the relation between parameters $B$ and $\beta$.

\noindent
{\it (i)} When $B > 1 {+} \beta$, or equivalently
\begin{equation}
\sigma > \frac{3}{\mu{-}2} {+} \xi (\mu{+}1)
\,,
\label{trans4}
\end{equation}
the discriminant in (\ref{trans3}) is always positive, the roots of characteristic equations are real and negative, thus the
second (quasi)critical point is a stable node for each $\tau_0 > \tau_{({\rm crit})}$.
We deal with the phase portrait, which contains a stable node $M_{(**)}$ and a saddle point $M_{(*)}$.

\noindent
{\it (ii)} When $B < 1 {+} \beta$  we face the reconstruction of the phase portraits with time. Indeed, one can fix two moments of time, $\tau^{(1)}_0$ and  $\tau^{(2)}_0$, defined as
$$
X_{(**)}(\tau^{(1)}_0) \equiv B-2+ 2\sqrt{1+\beta-B} \,,
$$
\begin{equation}
X_{(*)*}(\tau^{(2)}_0) \equiv B-2 - 2\sqrt{1+\beta-B}
\,,
\label{trans41}
\end{equation}
for which the discriminant in (\ref{trans3}) takes zero value. Since $X_{(**)}$ tends to zero with time, one obtains three subsequent situations:
first, when $\tau_{({\rm crit})}< \tau_0 < \tau^{(1)}_0$ the roots are real, i.e., $M_{(**)}$ is a node; second, when $\tau^{(1)}_{0}< \tau_0 < \tau^{(2)}_0$,
the roots are complex numbers, i.e., $M_{(**)}$ is a focus; third, when $\tau_{(2)}< \tau_0$ the
roots are real again, i.e., $M_{(**)}$ is a node. One can guarantee, that the nodes and focus are stable, if $\beta < B < 1+\beta$.

\noindent
{\it (iii)} If  $B < \beta$, evidently, there exists a moment $\tau^{(3)}_0$, when the decreasing quantity $X_{(**)}(\tau_0)< \beta$ becomes equal to the parameter $B$.
At this moment $\lambda_{1,2} = \pm i \sqrt{\beta {-}B}$, i.e., the focus transforms into a center.
It is possible, when
\begin{equation}
{-}1 < \sigma < 3\frac{[1{-}\xi(\mu{+}1)]}{(\mu{-}2)} \,, \quad \xi<\frac{1}{3}\,.
\label{trans6}
\end{equation}
Finally, at the moments $\tau^{(1)}_{0}$ and $\tau^{(2)}_0$ the discriminant in (\ref{trans3}) vanishes, i.e., the corresponding points should be classified as degenerated nodes.

\subsubsection{Phase portrait distortion and transformation}

Let us consider the plots in Fig.1(c),2(c),3(c),4(c),5(c),6(c), and 7(c), obtained numerically, from the point of view of {\it qualitative} analysis of the models with $\sigma>{-}1$.
In fact we take two typical phase portraits: for a node and a saddle point [Fig.8], and for a focus and saddle point [Fig.9], and provide the changes of two types. First, we move the (quasi)critical points horizontally, thus distorting the instantaneous phase portrait;  second, we replace the node by focus and vice-versa, thus modeling transitions in the instantaneous phase portraits with time.

\noindent
1. Figure1(c): {\it perpetually accelerated universe}.

\noindent
For the given values of the guiding parameters we have $\sigma < 3{+}4\xi$, i.e., the inequality $B < 1 {+} \beta$ is valid [see the discussion of Eq. (\ref{trans41}) above].
The zone $\tau < 10^0$ can be indicated as precritical, the function $\Pi(\tau)$, the DE pressure, decreases there, since the initial value $\Pi^{\prime}(1){=} {-}1$ is negative. Then, as it was
qualitatively predicted, the focus appears on the instantaneous phase portrait, so that the DE pressure enters the zone of damped quasioscillations.

\noindent
2. Figure 2(c): {\it periodic universe}.

\noindent
For this model, clearly, both inequalities given by (\ref{trans6}) are valid; in addition  $B < 1 {+} \beta$, thus it is a typical model with center [see the discussion od Eq. (\ref{trans6}) above]. The presence of
a center on the phase portrait explains the periodic graph of the DE pressure $\Pi(\tau)$ (the graph is presented in a logarithmic scale).

\noindent
3. Figure 3(c): {\it first model with one transition point}.

\noindent
There, clearly, the inequality (\ref{trans4}) is valid, i.e.,  $B > 1 {+} \beta$ and we deal with the absence of focuses. Instantaneous phase portraits contain stable nodes only, the graph for the DE pressure has no quasioscillations.

\noindent
4. Figure 4(c): {\it second model with one transition point}.

\noindent
5. Figure 5(c): {\it model with two transition points}.

\noindent
There, in both cases,  we deal with the situation similar to the one in Fig.1(c).

\noindent
6. Figure 6(c): {\it model with three transition points}.

\noindent
The situation is analogous to the case given by Fig.3(c).

\noindent
7. Figure 7(c): {\it quasiperiodic universe}.
Again we deal with the set of guiding parameters, for which  $B < 1 {+} \beta$, and thus the stable focus appears in the instantaneous phase portrait of the dynamic system;
quasioscillations of the DE pressure are typical for this model.

\subsection{Asymptotic behavior of the integral curves}

Speaking about nonrelativistic and ultrarelativistic models, we distinguish between
{\it initially} and {\it effectively} hot or cold states of the DM particles. When the parameters
$\lambda_{({\rm a})} \equiv \frac{m_{({\rm a})} c^2}{k_{({\rm B})} T{({\rm a})}}$ are
much bigger or much smaller than 1, we deal with initially nonrelativistic or ultrarelativistic DM,
respectively. In the course of the Universe expansion an initially cold DM can
become effectively hot (effectively ultrarelativistic), when the term $F_{({\rm
a})}(x) \to \infty$ [see (\ref{FF})]. Such terminology is advocated by the
fact, that at $F_{({\rm a})}(x) \to \infty$ one obtains
$<q^2> F_{({\rm a})}(x) >> 1$, and the averaged kinetic energy of a DM particle
becomes much bigger than the corresponding rest energy, even if initially at $t{=}t_0$ ($F(1){=}1$)
the inequality had the form $<q^2> << 1$. Analogously, when $F_{({\rm a})}(x) \to 0$, an
initially hot DM converts into the effectively nonrelativistic, because the
rest energy dominates over the kinetic energy of a DM particle. A new
interesting case can appear due to the Archimedean-type force, when $F_{({\rm a})}(x) \to const \neq
0$, despite the fact that $a(t) \to \infty$. In this case one deals
with the asymptotic fixation of the degree of relativism of the DM
particles instead of standard effective cooling. Let us consider these
asymptotic cases and analyze the model behavior for arbitrary values of the parameters $\lambda_{({\rm a})}$.

\subsubsection{Asymptotic  regime $F_{({\rm a})}(x) \to \infty$}

In the absence of the Archimedean-type force, i.e., at ${\cal V}_{({\rm a})}=0$, the
function $F_{({\rm a})}(x)=x^{-2}$ tends to zero asymptotically. When
${\cal V}_{({\rm a})} \neq 0$, the function $F_{({\rm a})}(x)$ can tend to infinity if the DE pressure
takes negative infinite value, $\Pi(x) \to - \infty$. In this asymptotic
regime, when $F_{({\rm a})}(x) \to \infty$, the integrals in (\ref{e(x)}),
(\ref{p(x)}), (\ref{key2}) can be easily calculated yielding
\begin{equation}
{\cal J}(x) {=} \frac{\Pi^{\prime}(x)}{x^3} \sum_{({\rm a})} E_{({\rm
a})} {\cal V}_{({\rm a})} I_{({\rm a})} e^{{\cal V}_{({\rm a})}
[\Pi(1){-}\Pi(x)]}
, \label{asy1}
\end{equation}
\begin{equation}
E(x) {=} 3P(x) {=} \frac{1}{x^3} \sum_{({\rm a})} E_{({\rm
a})} I_{({\rm a})} e^{{\cal V}_{({\rm a})}
[\Pi(1){-}\Pi(x)]}
, \label{asy2}
\end{equation}
where the parameter $I_{({\rm a})}$ is given by
\begin{equation}
I_{({\rm a})} \equiv 2 \lambda_{({\rm a})}^{{-}4}
[\lambda_{({\rm a})}^2 {+} 3\lambda_{({\rm a})} {+} 3] e^{{-}\lambda_{({\rm a})}} \,. \label{asy3}
\end{equation}
Let us indicate the maximal parameter from the set $\{ {\cal{V}}_{({\rm
a})}\}$ as $\nu$, the corresponding parameter $E_{({\rm a})}$ as $E$ and $I_{({\rm a})}$ as $I$. Then one obtains
that, when $\sigma > {-}1$ and $x \to \infty$, the leading order term in the left-hand side of the key equation
(\ref{key1}) is $3(1{+}\sigma) \Pi$ and the reduced equation for $\Pi(x)$ is
of the form
\begin{equation}
\Pi^{\prime}(x) = \frac{3(1+\sigma)}{\nu E I} x^3 \Pi(x) \exp{\{ \nu \Pi(x)\}} \,. \label{asy4}
\end{equation}
This means that at $\Pi \to {-} \infty$ the derivatives $\Pi^{\prime}$, $\Pi^{\prime
\prime}$, etc., asymptotically vanish, and the behavior of the DE pressure
$\Pi(x)$ is described by the function inverse to the integral exponent $Ei^{-1}(\Pi)$, where
$Ei(\Pi)=\int \frac{d\Pi}{\Pi} e^{-\nu \Pi}$. The decomposition of this function at $\Pi \to - \infty$ gives
the transcendent equation
\begin{equation}
\frac{1}{\Pi}\exp{\{-\nu \Pi(x)\}} = - \frac{3(1+\sigma)}{4 E I} x^4  \,. \label{asy5}
\end{equation}
When $\sigma = -1$, the asymptotic law is $\Pi(x) \to - \frac{4}{\nu}
\log{x}$ in accordance with the exact solution of the anti-Gaussian type,
studied in the first part of the work \cite{antigauss}.
The asymptotic behavior of the DE energy density is given by the formula $\rho(x) \simeq \sigma \Pi$ [see (\ref{simplest0})]
and depends essentially on the sign of the parameter $\sigma$: it remains positive at $\Pi(x) \to {-} \infty$, when
${-}1< \sigma <0$.
Concerning the DM energy density $E(x)$ and DE pressure $P(x)$, these
quantities grow exponentially; we do not consider the Hubble function $H(x)$
and the scale factor $a(t)$ for such models, since there is no observational
data, which could confirm such behavior.

\subsubsection{Asymptotic  regime $F_{({\rm a})}(x) \to 0$}

This asymptotic regime can be realized, when the DE pressure takes arbitrary positive or finite negative
value at $x \to \infty$. The asymptotic behavior of this type relates to the vanishing source term
$$
{\cal J}(x) \to  \frac{3 \Pi^{\prime}}{x^4} {\cal N}
\exp{\{2 {\cal V}_{({\rm a})} [\Pi(1){-}\Pi(x)]\}} \,,
$$
\begin{equation}
{\cal N} {=}  \sum_{({\rm a})} N_{({\rm a})} k_{({\rm B})} T_{({\rm a})} {\cal V}_{({\rm a})}
\frac{K_3(\lambda_{({\rm a})})}{K_2(\lambda_{({\rm a})})} \,,
\label{2asy1}
\end{equation}
and vanishing DM energy density
\begin{equation}
E(x) = \frac{1}{x^3} \sum_{({\rm a})} N_{({\rm a})} m_{({\rm a})} \,.
\label{2asy2}
\end{equation}
Thus, the DE pressure satisfies now the Euler equation
\begin{equation}
\xi x^2 \Pi^{\prime \prime}(x) + x \Pi^{\prime}(x) \left(4 \xi +
\sigma \right) + 3 (1+\sigma)\Pi +
3 \rho_0 = 0 \,,
\label{2asy3}
\end{equation}
which is studied in detail in Sec.III A of the first part of our work \cite{antigauss}.
In particular, in Sec.III A 3, we have shown that there exist quasiperiodic oscillations
near the asymptote $\Pi(x) = -\frac{\rho_0}{1+\sigma}$, which are manifestly
presented in Fig.1(c), Fig.4(c), Fig.5(c), Fig.7(c). The aperiodic
solutions studied in Sec.III A 1, and Sec.III A 2 \cite{antigauss}
clearly appear in Figs.3(c) and 6(c).

\subsubsection{Asymptotic  regime $F_{({\rm a})}(x) \to const \neq 0$}

Let us suppose for simplicity that ${\cal V}_{({\rm a})} {=} \nu$ for all types of the DM particles, and let the corresponding function
$F(x)$ tend to a nonvanishing constant value $F_{\infty}$, when $x \to
\infty$. This means that asymptotic behavior of the DE pressure is the
following
\begin{equation}
\Pi(x) \to \Pi(1) - \frac{1}{\nu} \log{x} - \frac{1}{2\nu} \log{F_{\infty}} \,. \label{3asy1}
\end{equation}
Then the initial data require that $F_{\infty}=1$, and the source term (\ref{key2}) takes asymptotically
the following form
\begin{equation}
{\cal J}(x) \to {-} \frac{3\nu}{x^3} \sum_{({\rm a})} N_{({\rm a})} k_{({\rm B})} T_{({\rm a})} \,,
\label{3asy2}
\end{equation}
i.e., it decreases as $x^{-3}$ at $x \to \infty$.
The left-hand side of Eq. (\ref{key1}) vanishes after the substitution of (\ref{3asy1}), if $\sigma= {-}1$
and $3\rho_0 \nu = 3\xi {-}1$, i.e., the asymptotic regime exists at special conditions only.
The DE and DM energy densities behave in this case  as
$$
\rho(x) \to \frac{1}{\nu} \log{x} \,,
$$
\begin{equation}
E(x) \to  \frac{1}{x^3} \sum_{({\rm a})} N_{({\rm a})} k_{({\rm B})} T_{({\rm a})}
\left[\lambda_{({\rm a})} \frac{K_3(\lambda_{({\rm a})})}{K_2(\lambda_{({\rm a})})} {-}1 \right]
\,.
\label{3asy3}
\end{equation}
The Hubble function, scale factor and acceleration parameters are given by the following formulas
$$
H(x) \to \sqrt{\frac{8\pi G}{3 \nu} \log{x}}  \,, \quad a(t) \to a(t_0) \exp{\left\{\frac{2\pi G}{3\nu}t^2 \right\}} \,,
$$
\begin{equation}
-q(t) \to 1 +
\frac{3\nu}{4\pi G t^2}
\,,
\label{3asy4}
\end{equation}
which, clearly, describe the anti-Gaussian solution studied in the first
part of the work \cite{antigauss}.

\section{Discussion}

The Archimedean-type model can describe qualitatively two cornerstones  of
the Universe's evolution: the inflation in the early Universe and the late-time accelerated expansion.
Also, this model offers a variant explanation of the so-called coincidence problem, and it gives some motivation for the specific division
of the Universe's history into epochs and eras. Let us consider these questions in more detail.

\subsection{Inflation in the early Universe}

When $\tau \leq 1$ Fig.1(a), 2(a),3(a),4(a),5(a),6(a), and 7(a) display the first inflationary-type
epoch: one can reveal a quick growth of the DE energy density $\rho(\tau)$.
During this period of the Universe evolution the scale factor $a(t)$
increases essentially: in order to visualize this inflationary effect we presented additionally the fragment of the plot
$a(t)$ in the natural scale $t$ instead of the logarithmic one [see Fig.11(a)]. The steepness of the plot of $a(t)$ is predetermined by the
combination of the guiding parameters; there are a lot of possibilities to fit the observed curve by the curve predicted in the framework
of the Archimedean-type model.
In the context of expansion the growth of the DE energy density $\rho(\tau)$ in the early Universe
relates to the heating of the DE component of the dark fluid. Clearly [compare, e.g., the panels (a) and
(b) in Figs.1-7] the sum $\rho{+}E$ remains non-negative for the whole interval of
time. This feature provides the $H^2$ to be non-negative, which guarantees that $H$ is a real function.
At the same time the DM energy density $E(\tau)$ [see the panels (b)] decreases monotonically; nevertheless,
the rate of its effective cooling differs essentially from the rate given by the standard powerlaw function ($1/a^4$
for the ultrarelativistic DM or $1/a^3$ for the cold DM). Principally new
detail can be found in Fig.4(d) and Fig.5(d):
the DM pressure is described by a nonmonotonic function; it grows quickly, reaches a
maximum, and then decreases rapidly.

\begin{figure}
\centerline{\includegraphics{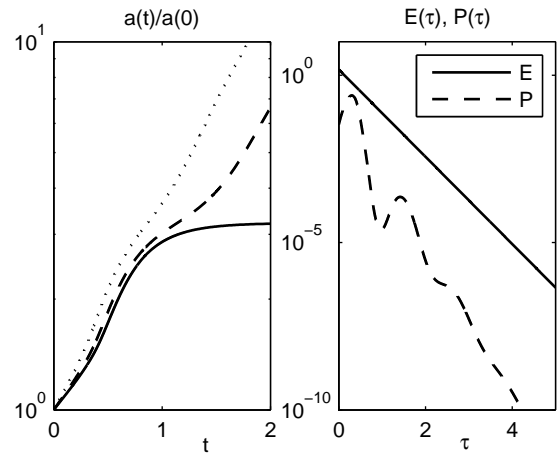}}
\caption {{\small In the left panel a fragment of the plot is presented, which illustrates the inflationary-type growth of the
scale factor $a(t)$ for the early Universe, given in terms of cosmological time $t$. In the right panel one can see a typical late-time behavior of
$E(\tau)$ and $P(\tau)$; this rescaled fragment illustrates the features of the DM evolution at $\tau >1$, which cannot be recognized on the plots given by Fig.1(b,d)-Fig.7(b,d).
}}
\end{figure}

\subsection{Late-time accelerated expansion}

When $\tau \to \infty$, the curves ${-}q(\tau)$, which present the evolution of the acceleration parameter in  Fig.1(f),2(f),3(f),4(f),5(f),6(f), and 7(f)
tend to the horizontal asymptote, the value ${-}q(\infty)$ being positive. Taking into account the behavior of the plots
of the Hubble functions displayed in Fig.1(e), Fig.3(e)-Fig.7(e), one can conclude that
$H(\tau \to \infty) \to H_{\infty}$, i.e., the Hubble function also tends to the positive constant value,
when $\tau \to \infty$.
This means, first, that a typical behavior of the Universe with Archimedean-type
interaction between DE and DM is characterized by the late-time accelerated expansion; second, that
asymptotically the Archimedean-type model converts into the quasi-de Sitter one.

The DM state functions $E(\tau)$ and $P(\tau)$ decrease rapidly, and for $\tau > 1$ we should change the scale in order to visualize the behavior
of these functions. A typical late-time behavior of  $E(\tau)$ and $P(\tau)$
is presented in the right panel of Fig.11. In order to describe the effective equation of state of dark matter in the framework of the Archimedean-type model,
we calculated numerically the ratio $w(\tau)\equiv \frac{P(\tau)}{E(\tau)}$; the plot of the function $w(\tau)$ is presented in the left panel of Fig.12.
Clearly, this function is constrained: $0<w(\tau)<1$; maximal value of this ratio is about $1/3$ in the first epoch of the Universe's evolution, when the DM can be considered
as an effectively ultrarelativistic substrate. In the early Universe the  DM state evolves quasiperiodically, i.e., the eras with effective cooling were changed by the eras
of effective heating. Starting from $\tau \simeq 4$ the function $w(\tau)$ tends to zero linearly in $\tau {=} \log{\frac{a(t)}{a(t_0)}}$, i.e., the DM behaves effectively
as a cold gas (dust).

\begin{figure}
\centerline{\includegraphics{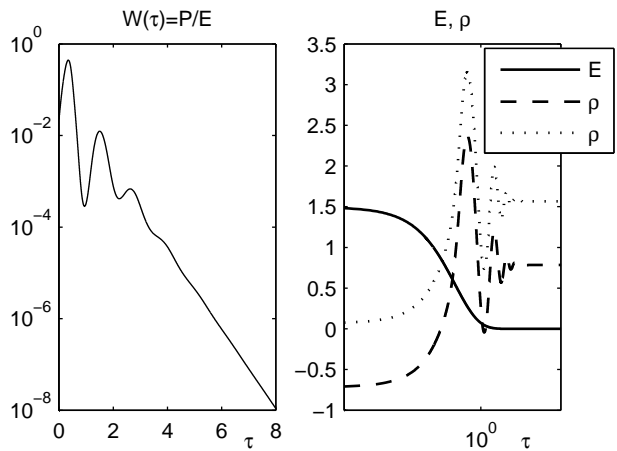}}
\caption {{\small The left panel contains the plot of the ratio $w(\tau){=}\frac{P(\tau)}{E(\tau)}$, which presents the effective equation of state for the DM
component of the dark fluid. Maximal value of this ratio is about $1/3$ in the first epoch of the Universe's evolution, when the DM can be considered
as an effectively ultrarelativistic substrate. In the early Universe the  DM state evolves quasiperiodically; starting from $\tau \simeq 4$ the function $w(\tau)$ tends
to zero linearly in $\tau$, i.e., the DM behaves effectively as a cold gas (dust).
The right panel displays the cross-points of the $\rho$ and $E$ plots. This example illustrates the so-called coincidence problem: why
the energy densities of the DE and DM components of the dark fluid are of the same order today ($72\%$ and $23\%$, respectively).
}}
\end{figure}

\subsection{Coincidence problem}

Although the DE and DM components of the dark fluid evolve at different rates throughout the history of the Universe,
their magnitudes are of the same order today ($72\%$ and $23\%$, respectively). This is known as the coincidence problem
\cite{coi1}-\cite{coi4}. The Archimedean-type model could help us to make a step toward solving this problem. In the right panel of Fig.12 we placed the example
of the plots describing the DE and DM evolution with time $\tau$. The guiding parameter $\rho^{*}_0$ defines the asymptotical ratio between $\rho$ and $E$; this ratio can
in principle be chosen so that $\frac{\rho}{E}\to \frac{72}{23}$. Other guiding parameters define how many cross-points of the $\rho$ and $E$ plots can exist. For instance,
in Fig.12 there are examples with one and three cross-points. Using the guiding parameters we can remove one of the cross-points away from the point $\tau \simeq 1$, but
we hope to discuss in detail the fitting problem in a special work.

\subsection{On the partition of the Universe's history into epochs and eras and its multi-inflationary behavior}

There are a few versions of dividing the history of the Universe into self-sufficient parts that are interesting from a physical point of view. For instance, one can consider the
Hubble function $H(t)$, find its zeros and extrema and separate the admissible time interval in line with them (see, e.g., \cite{Izq}. We attached such a partition to the function ${-}q(t)$,
i.e., based the classification of models on the zeros and extrema of the acceleration parameter. This seems to be motivated, since we are interested in picking out the epochs of
accelerated and decelerated expansion of the Universe, thus the transition points appeared as points, in which the function $-q(t)$ changes the sign. We have shown above that there
are a lot of models in which the Universe's history is multistage. In particular, the Universe's evolution can be quasiperiodic, and one can use the term {\it multi-inflationary} evolution,
the first inflation being the sharpest, others being more and more smoothed.  Following this line, we 
divide the epochs into eras by using the maximums and minimums of this function: this is equivalent to the method of statefinders proposed in \cite{statefinder}. We hope to consider such
fine details of the Universe's history partition in a special work.

\subsection{Conclusions}

The model of Archimedean-type coupling between dark energy and dark matter makes it possible to explain the principal cornerstones of the
Universe's evolution: the early-time inflationary expansion, the late-time accelerated expansion, and the coincidence phenomenon.

The model of Archimedean-type coupling between dark energy and dark matter possesses a wide set of guiding parameters suitable for fitting of the model predictions
to the observational data; we hope to devote a special work to this important question.

The model of Archimedean-type coupling offers a natural approach to the partition of the Universe's history into epochs and eras using the number of transition
points and number of extrema of the function ${-}q(t)$, the acceleration parameter of the Universe's evolution. The model allows us to speak about multistage
evolution and about the multi-inflationary behavior of the Universe.

\section{Acknowledgments}
The authors are grateful to Professor W. Zimdahl for fruitful discussions, comments and advice.
This work was partially supported by the Russian Foundation for Basic Research
(Grants No. 08-02-00325-a and 09-05-99015) and by Federal Targeted Programme, Scientific and Scientific-Pedagogical Personnel of the Innovative Russia
(Grants No. 16.740.11.0185 and  14.740.11.0407).

\end{document}